# Optimizing the signal-to-noise ratio of biphoton distribution measurements

Matthew Reichert,[*] Hugo Defienne, and Jason W. Fleischer[†]

*Department of Electrical Engineering, Princeton University, Princeton, New Jersey 08544, USA*

Single-photon-sensitive cameras can now be used as massively parallel coincidence counters for entangled photon pairs. This enables measurement of biphoton joint probability distributions with orders-of-magnitude greater dimensionality and faster acquisition speeds than traditional raster scanning of point detectors; to date, however, there has been no general formula available to optimize data collection. Here we analyze the dependence of such measurements on count rate, detector noise properties, and threshold levels. We derive expressions for the biphoton joint probability distribution and its signal-to-noise ratio (SNR), valid beyond the low-count regime up to detector saturation. The analysis gives operating parameters for global optimum SNR that may be specified prior to measurement. We find excellent agreement with experimental measurements within the range of validity, and discuss discrepancies with the theoretical model for high thresholds. This work enables optimized measurement of the biphoton joint probability distribution in high-dimensional joint Hilbert spaces.

## I. INTRODUCTION

Quantum states of light—such as entangled photon pairs (biphotons)—offer substantial promise over classical light, including enhanced spatial resolution, phase sensitivity, and signal-to-noise ratio [1-3]. They also hold great potential in quantum metrology, with possible improvements in gravitational wave detection [4], biology [5], and microscopy [1]. Increasingly, these fields are moving towards higher-dimensional entanglement, as it offers greater channel capacity [2,6-9], security [2,6-9], and computational speed [10,11].

Garnering these advantages requires detecting both photons in coincidence. This is typically performed with two single photon-counting modules (SPCMs); as these modules have no spatial resolution, both must be scanned over each dimension of the joint Hilbert space. In an imaging configuration, for example, measuring photon pairs entangled in transverse position requires scanning each detector over a 2D plane. The number of required measurements scales quadratically with the number of modes, making high-dimensional entangled systems prohibitively time consuming to characterize and inaccessible in practice. Furthermore, coincidence measurements of biphotons are typically performed in the low-count regime, where the count rate itself may be assumed to be proportional to the biphoton joint probability distribution. Operating at a substantially higher count rate can yield drastic improvements in measurement speed and signal-to-noise ratio (SNR) [12], but direct proportionality breaks down as the count rate increases (accidental coincidences between photons from different pairs become substantial or even dominant). This breakdown complicates the relationship between the measured detection counts and the true joint probability distribution, making interpretation of the results far from straightforward.

The advent of single-photon-sensitive cameras, such as intensified CCD (ICCD) and electron-multiplying CCD (EMCCD) cameras, has made rapid characterization of the spatially entangled photon pairs feasible [12-21]. We have recently developed a method of parallelizing such measurements using an EMCCD camera [19]. Each pixel is treated as a single-photon counter, with coincidences between all pixels measured simultaneously. Using only measured data, we have shown how to account completely for genuine and accidental coincidences. For a megapixel camera, the massively parallel apparatus allows for precise measurement of the biphoton joint probability distribution within Hilbert spaces of up to $10^{12}$ dimensions. Such measurements are impractical with traditional scanning (or compressed sensing) methods.

The goal of this work is to provide a prescription for optimizing the measurement of the biphoton joint probability distribution. Prior work has examined maximizing the visibility of the genuine biphoton coincidences relative to the accidentals background [22]. There, the authors found an optimum visibility when the count rate from photons is equal to that from electronic noise events but noted that the SNR could be improved by increasing the count rate. Similarly, Lantz *et al.* found that the SNR is improved for higher count rates [12], provided that measurements remain within the low count-rate regime. Indeed, if the background can be identified and removed, only the fluctuations in the background limit the quality of the result.

Here, we develop a general model for the SNR of measurements of the biphoton joint probability distribution that is valid for arbitrary count rates, up to saturation of the detector. Our model is based on binary detection systems and accounts completely for multiple photons and their number distribution. The SNR is given in terms of the singles count rate and detector noise properties and allows optimization of any part of the distribution function, including and especially coincidence measurements of entangled photon pairs.

We apply this model to massively parallel coincidence counting with EMCCD cameras [19] and compare to experimental measurements of spatially entangled biphotons. We operate the camera in photon-counting mode and consider detection as a function of gray level threshold. For low threshold, EMCCDs are well approximated as binary detection devices [23,24]. For higher thresholds, this approximation breaks down. Experimentally, we explore both regions, with a focus on the validity of the binary model and its impact on the SNR. Note that while we consider spatial entanglement here, the analysis applies to other degrees of freedom as well, such as frequency or orbital angular momentum, with appropriate projection onto camera pixels.

## II. THEORY

The (pure) quantum state of entangled photon pairs may be defined by

$$|\Psi\rangle = \iint \psi(\boldsymbol{\rho}_1, \boldsymbol{\rho}_2)|\boldsymbol{\rho}_1\rangle|\boldsymbol{\rho}_2\rangle d^2\boldsymbol{\rho}_1 d^2\boldsymbol{\rho}_2, \quad (1)$$

where $\psi(\boldsymbol{\rho}_1, \boldsymbol{\rho}_2)$ is the transverse biphoton wave function and $|\boldsymbol{\rho}_i\rangle$ are states of the transverse position with $\boldsymbol{\rho} = x\hat{\boldsymbol{x}} + y\hat{\boldsymbol{y}}$. We want to measure the biphoton probability distribution $|\psi(\boldsymbol{\rho}_1, \boldsymbol{\rho}_2)|^2$ using an EMCCD camera. We thus have a discretized distribution

$$\Gamma_{ij} = \iint_{-w/2}^{w/2} |\psi(\boldsymbol{\rho}_1 - \boldsymbol{\rho}_i, \boldsymbol{\rho}_2 - \boldsymbol{\rho}_j)|^2 d^2\boldsymbol{\rho}_1 d^2\boldsymbol{\rho}_2, \quad (2)$$

where $w$ is the width of the square pixels centered at positions $(\boldsymbol{\rho}_i, \boldsymbol{\rho}_j)$. The marginal distribution is

$$\Gamma_i = \sum_j \Gamma_{ij}$$
$$= \int_{-w/2}^{w/2} \left[ \int |\psi(\boldsymbol{\rho}_1 - \boldsymbol{\rho}_i, \boldsymbol{\rho}_2)|^2 \, d^2\boldsymbol{\rho}_2 \right] d^2\boldsymbol{\rho}_1, \quad (3)$$

which is proportional to the irradiance.

In general, there are two possible cases: (1) photons from pairs are deterministically separated to different detector arrays (or different regions of a single array), and (2) photons are all sent to a single detector array. The principle difference between them is that in (2), both photons from a single pair may hit the same pixel,. Case (1) is only possible for distinguishable particles, where some degree of freedom uniquely identifies which photon is which, e.g., polarization, frequency, etc. Here we present equations for case (1) explicitly. To convert to case (2), the substitutions $\Gamma_i \to 2\Gamma_i - \Gamma_{ii}$ and $\Gamma_{ij} \to 2\Gamma_{ij}$ should be made throughout. In addition, to simplify the notation, we omit factors of the detector quantum efficiency $\eta$; to account for it, one need only make the substitutions [25]

$$\begin{aligned}\Gamma_i &\to \eta \Gamma_i \\ \Gamma_{ij} &\to \eta^2 \Gamma_{ij}\end{aligned} \quad (4)$$

### A. Coincidence count distributions

The singles count probability at pixel $i$ is given by

$$\langle C_i \rangle = \sum_m P_m \left( \mu_{i|m} + p_{el} \mu_{\bar{i}|m} \right) \quad (5)$$

where $P_m$ is the probability distribution for the number of generated pairs $m$ and $p_{el}$ is the electronic count probability of the detector (dark counts, CIC, etc.). The first term describes the probability of counts due to photons, while the second describes counts due to electronic noise in the absence of photons. The factor $\mu_{\bar{i}|m}$ is the conditional probability, given $m$ photon pairs, that no photons are detected in pixel $i$ (indicated by the barred $\bar{i}$), which is related to the marginal distribution by

$$\mu_{\bar{i}|m} = (1 - \Gamma_i)^m. \quad (6)$$

$\mu_{i|m}$ is the conditional probability that *at least* one photon is detected in pixel $i$. Because the two conditionals sum to unity, they are related by

$$\mu_{i|m} = 1 - \mu_{\bar{i}|m} \quad (7)$$

In a similar fashion, the coincidence count probability between pixels $i$ and $j$ may be written

$$\langle C_{ij} \rangle = \sum_m P_m [\mu_{ij|m} + p_{el}(\mu_{i\bar{j}|m} + \mu_{\bar{i}j|m}) + p_{el}^2 \mu_{\bar{i}\bar{j}|m}]. \quad (8)$$

The first term represents coincidences between two photons, the second between one photon and one electronic noise event, and the third between two noise events. The sum of the $\mu$'s gives unity: $\mu_{ij|m} + \mu_{i\bar{j}|m} + \mu_{\bar{i}j|m} + \mu_{\bar{i}\bar{j}|m} = 1$.

As before, to find the full expression for $\langle C_{ij} \rangle$, it is easiest to consider the zero-photon case first. Coincidences between two electronic noise events depend on no photon detection in either pixel $i$ or $j$, so that

$$\mu_{\bar{i}\bar{j}|m} = \left(1 - \Gamma_i - \Gamma_j + \Gamma_{ij}\right)^m. \quad (9)$$

The coincidence counts between photons and electronic noise requires *at least* one photon detection in one pixel and zero in the other. This is given by the probability that no photons are detected in one pixel, i.e., $\mu_{\bar{j}|m}$, minus the probability that no photons are detected in either pixel, $\mu_{\bar{i}\bar{j}|m}$, that is

$$\mu_{i\bar{j}|m} = \mu_{\bar{j}|m} - \mu_{\bar{i}\bar{j}|m}. \quad (10)$$

where $\bar{j}$ indicates that mode $j$ is unoccupied, and vice-versa for $\mu_{\bar{i}j|m}$. The probability that *at least* one photon is detected in each pixel $i$ and $j$ is then

$$\mu_{ij|m} = 1 - \mu_{\bar{i}|m} - \mu_{\bar{j}|m} + \mu_{\bar{i}\bar{j}|m} \quad (11)$$

Equations (5) and (8) have simple analytic form if the number distribution of pairs is Poissonian. In this case, $P_m = \bar{m}^m e^{-\bar{m}}/m!$, where $\bar{m}$ is the mean number of pairs emitted within exposure time $\tau_e$ [26,27], and we have

$$\langle C_i \rangle = 1 - (1 - p_{el})e^{-\bar{m}\Gamma_i} \quad (12)$$

and

$$\langle C_{ij}\rangle = 1 - (1-p_{el})\left(e^{-\bar{m}\Gamma_i} + e^{-\bar{m}\Gamma_j}\right) \\ + (1-p_{el})^2 e^{-\bar{m}(\Gamma_i+\Gamma_j-\Gamma_{ij})} \quad (13)$$

Notice that both $\langle C_i\rangle$ and $\langle C_j\rangle$ appear within $\langle C_{ij}\rangle$. Using Eq. , we can rewrite Eq. (13) as

$$\langle C_{ij}\rangle = \langle C_i\rangle + \langle C_j\rangle - 1 \\ + (1-\langle C_i\rangle)(1-\langle C_j\rangle)e^{2\bar{m}\eta^2\Gamma_{ij}} \quad (14)$$

where we have re-introduced the quantum efficiency $\eta$. Solving for $\Gamma_{ij}$ gives

$$\Gamma_{ij} = \alpha \ln\left[1 + \frac{\langle C_{ij}\rangle - \langle C_i\rangle\langle C_j\rangle}{(1-\langle C_i\rangle)(1-\langle C_j\rangle)}\right] \quad (15)$$

where $\alpha = 1/(\bar{m}\eta^2)$ [$\alpha = 1/(2\bar{m}\eta^2)$ for case (2)]. Therefore, to within a constant scaling factor, only the mean coincidence- and singles-count probabilities are necessary to uniquely extract the joint probability distribution.

### B. Signal-to-noise ratio

We want to relate the signal that we measure to the mean count rates. To do so, we write $\Gamma_{ij} = \Gamma_i \Gamma_{j|i}$ and use Eq. (15) to express $\Gamma_i$ in terms of the singles-count probability:

$$\Gamma_{ij} = -\alpha\eta\Gamma_{j|i} \ln\left(\frac{1-\langle C_i\rangle}{1-p_{el}}\right). \quad (16)$$

The estimator of the joint probability distribution has standard deviation

$$\sigma_{\Gamma_{ij}} = \sigma_{\langle C_{ij}\rangle} \frac{\partial \Gamma_{ij}}{\partial \langle C_{ij}\rangle} \quad (17)$$

where

$$\sigma_{\langle C_{ij}\rangle} = \sqrt{\frac{1}{N}\langle C_i\rangle(1-\langle C_i\rangle)\langle C_j\rangle(1-\langle C_j\rangle)} \quad (18)$$

and

$$\frac{\partial \Gamma_{ij}}{\partial \langle C_{ij}\rangle} = \alpha \frac{1}{1-\langle C_i\rangle - \langle C_j\rangle + \langle C_{ij}\rangle} \quad (19)$$

giving

$$\sigma_{\Gamma_{ij}} = \frac{\alpha}{\sqrt{N}} \frac{\sqrt{\langle C_i\rangle(1-\langle C_i\rangle)\langle C_j\rangle(1-\langle C_j\rangle)}}{1-\langle C_i\rangle - \langle C_j\rangle + \langle C_{ij}\rangle} \quad (20)$$

We define the noise where $\Gamma_{ij} = 0$, i.e., where $\langle C_{ij}\rangle = \langle C_i\rangle\langle C_j\rangle$. For uniform illumination, $\langle C_i\rangle = \langle C_j\rangle = \langle C\rangle$, the noise becomes

$$\sigma_{\Gamma_{ij}} = \frac{\alpha}{\sqrt{N}} \frac{\langle C\rangle}{1-\langle C\rangle} \quad (21)$$

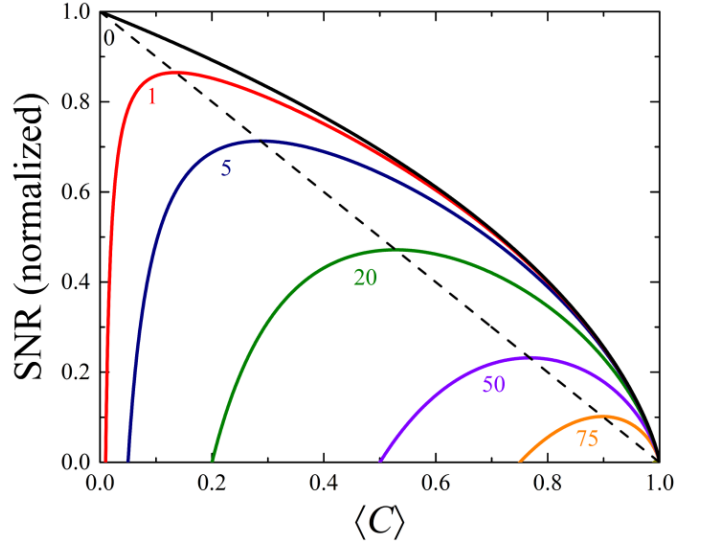

Fig. 1 Normalized signal-to-noise ratio [SNR/$(\eta\Gamma_{j|i}\sqrt{N})$] versus mean count rate (solid curves) plotted for several values of $p_{el}$ (indicated by the numbers below each curve, in %). Dashed line shows the trend of the maximum SNR of the $p_{el}$, given by $1 - \langle C\rangle$.

(for non-uniform illumination, see Appendix A). The SNR is given by the ratio of Eq. (16) over Eq. (21):

$$\text{SNR} = \eta\Gamma_{j|i}\sqrt{N}\frac{\langle C\rangle - 1}{\langle C\rangle}\ln\left(\frac{1-\langle C\rangle}{1-p_{el}}\right) \quad (22)$$

In the low-count-rate limit, this equation reduces to the formula provided in [12].

Equation (22) relates the quality of measurements of the biphoton joint probability distribution to experimental parameters. These are either parameters set by the detection system—the quantum efficiency $\eta$ and electronic noise probability $p_{el}$—or set by the user—the mean count rate $\langle C\rangle$ and number of frames $N$. Figure 1 shows the normalized SNR (SNR divided by $\eta\Gamma_{j|i}\sqrt{N}$) versus $\langle C\rangle$ for several different values of $p_{el}$. In the limit of low electronic noise ($p_{el} \to 0$), the SNR is maximized for low count rate $\langle C\rangle \to 0$, since the only coincidence counts are those between entangled photon pairs. Increasing the count rate adds accidental coincidences between photons from different pairs, which contribute noise and reduce the SNR. When $p_{el}$ is nonzero, electronic noise dominates at low count rates, and the SNR increases with $\langle C\rangle$ until it reaches a maximum and turns back over. For high count rates, the number of accidentals grows more rapidly than those from entangled pairs, and the SNR $\to 0$ as $\langle C\rangle \to 1$.

The optimum count rate, i.e., the one that maximizes the SNR, is:

$$\langle C\rangle_{\text{opt}} = 1 + W\left(\frac{p_{el}-1}{e}\right) \quad (23)$$

where $W$ is the Lambert-W function [28]. $\langle C\rangle_{\text{opt}}$ depends only the noise characteristics of the detector; by identifying $p_{el}$, the mean count rate may be set by adjusting the pump power or exposure time. The corresponding maximum achievable SNR [peaks of curves in Fig. 1] falls off with more electronic noise. Remarkably, however, its falloff is quite slow; a relatively high $p_{el}$ of 0.2 yields a reduction in the maximum SNR of only 50 % from when $p_{el} = 0$.

Table I. Parameters for Andor iXon Ultra 897 EMCCD camera based on fit of histogram shown in Fig. 3 with Eqs. (25)-(28). EMCCD was set to readout rate of 17 MHz, 0.3 μs vertical shift time, vertical clock voltage set +4 V above default.

| Parameter | Value |
| --- | --- |
| $x_{el}/x$ | 12 |
| $g$ | 1000 |
| $r$ | 536 [29] |
| $p_c$ | 0.012971 |
| $\mu_{read}$ | 167.1035 gl |
| $\sigma_{read}$ | 18.379 gl |
| $p_{par}$ | $6.03 \times 10^{-3}$ |
| $p_{ser}$ | $5.32 \times 10^{-5}$ |

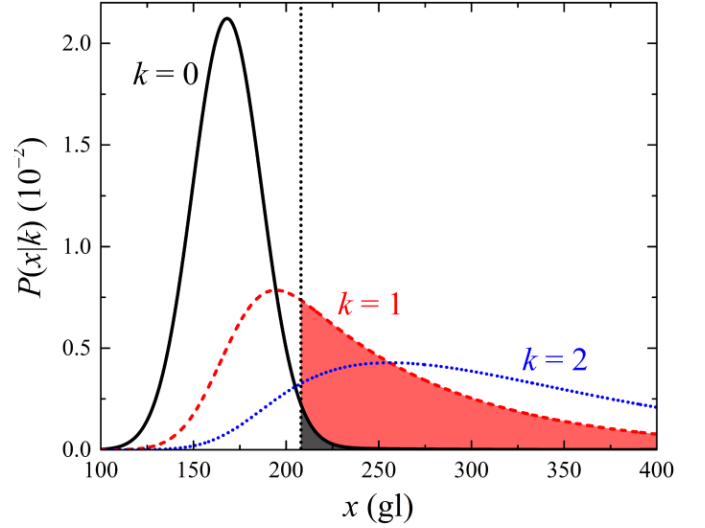

Fig. 2 Typical conditional probability distributions of gray-level (gl) outputs from EMCCD camera given (solid black) zero, (dashed red) one, and (dotted blue) two input photoelectron. Plots are based on Eqs. (25)-(28), with parameters in Table I. Vertical dotted line indicates typical threshold level of $T = 210$ gl. Shaded regions represent $P(x > T|k)$, the areas of which gives $p_{el}$ and $\eta_{EMCCD}$ for $k = 0$ and 1, respectively.

### C. Electron-Multiplying CCD Camera

Electron-multiplying charge-coupled-device (EMCCD) cameras are massively parallel single-photon-sensitive devices capable of measuring high-dimensional biphoton joint probability distributions [19]. If the camera is operated in photon-counting mode, where the gray levels above a thresholded value are registered as "clicks" and set to 1 while those below threshold are set to zero, then the probability of a gray level above threshold is

$$P(x > T|k) = \sum_{x>T}^{\infty} P(x|k) \quad (24)$$

where $k$ is the number of photoelectrons generated by the detector. This conditional probability distribution depends on the gain and noise properties of the EMCCD, and has been studied extensively [23,24]. In the following, we provide a summary of the principal contributions.

Photons incident on the camera are absorbed to create photoelectrons with quantum efficiency $\eta$. The electron-multiplying gain then amplifies the number $k$ of electrons stochastically, producing a random number of electrons at the output $x_{el}$, with conditional probability distribution $P(x_{el}|k)$. Photoelectrons at the input of the multiplication register produce an output number $x_{el}$ of electrons with conditional probability [23,24,30]

$$P_{gain}(x_{el}|n) = \frac{x_{el}^{n-1} e^{-\frac{x_{el}}{g}}}{g^n (n-1)!}, \quad (25)$$

where $g = (1 + p_c)^r$ is the mean gain, where $p_c$ is the multiplication probability in each of the $r$ elements in the multiplication register. Finally, an analog-to-digital converter produces a gray level value $x$ proportional to the number of electrons.

There are several processes that result in noise independent of the presence of photoelectrons. Readout noise yields a Gaussian distribution with mean $\mu_{read}$ and standard deviation $\sigma_{read}$;

$$P_{read}(x) = \frac{1}{\sqrt{2\pi}\sigma_{read}} e^{-\frac{(x-\mu_{read})^2}{2\sigma_{read}^2}}. \quad (26)$$

In addition, there are two noise processes that depend on the gain. First, there is a small probability $p_{par}$ that a spurious electron will be generated at the input of the multiplication register. This is predominantly due to clock-induced charge (CIC), as thermal dark counts are comparably negligible at low operating temperatures and short exposure times [24]. As this electron experiences the same gain as the photo-generated electrons, it results in a probability of electrons at the output of the multiplication

$$P_{par}(x_{el}) = p_{par} P_{gain}(x_{el}|1) = p_{par} \frac{e^{-\frac{x_{el}}{g}}}{g}. \quad (27)$$

Second, there is a small probability $p_{ser}$ that a spurious electron will be generated at each multiplication register cell, which is then amplified by the remaining registers. This results in an output probability

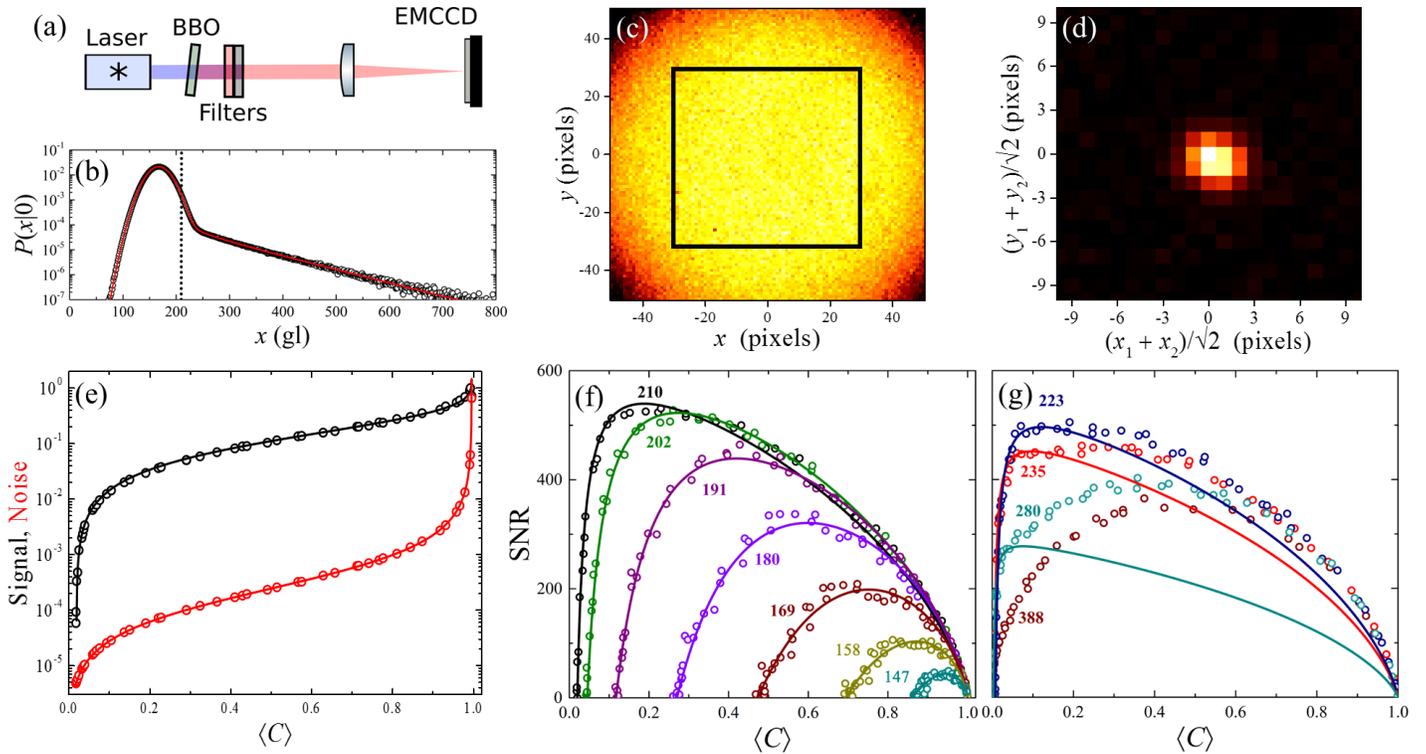

Fig. 3 Experimental measurement of SNR of entangled photon pairs. (a) Experimental schematic. A 400 nm laser diode pumps a BBO crystal and near-degenerate down-converted photons are filtered, and the far field projected onto an EMCCD camera (Andor, iXon Ultra 897). (b) Conditional probability distribution of gray-level output given zero input photoelectrons. (Black open circles) measured histogram of gray levels from ~$10^5$ 101×101 pixel frames collected with the shutter closed at 5 ms exposure time. (Red curve) Fit of Eqs. (25)-(28) to determine EMCCD properties given in Table I. Dotted vertical line shows threshold at $T = 210$ gl. The EMCCD measures (c) the irradiance distribution and (d) $\Gamma_{ij}$, shown projected onto the sum coordinates. A region of uniform irradiance, indicated by the central black boxed region in (c), is selected for SNR measurements. The signal and noise are taken as the area of the anti-correlation peak in (d) and the standard deviation of the fluctuations of the background far from the peak. Measurements (open circles) are repeated for many values of $\langle C \rangle$, and (e) the signal (black) and noise (red) are calculated and fit (solid curves) to theory. (f) Their ratio is taken to determine the SNR. Black dataset in (f) corresponds to signal and noise in (e). Also in (f), many lower thresholds were applied to the gray scale images, resulting in increased $p_{el}$ and $\eta_{EMCCD}$, all showing agreement with theory. Numbers next to each curve indicate threshold level (see Fig. 2 and Table I). (g) We observe disagreement that worsens for higher thresholds, where the approximations in treating the EMCCD as an SPCM break down.

$$P_{ser}(x_{el}) = p_{ser} \sum_{l=1}^{r} \frac{e^{-\frac{x_{el}}{(1+p_c)^{r-l}}}}{(1+p_c)^{(r-l)}}. \quad (28)$$

Both Eqs. (27) and (28) are valid only for $x_{el} > 0$; their value at $x = 0$ is determined by $1 - P_i(x_{el} > 0)$. Finally, the total $P(x|k)$ is given by the convolution of Eqs. (25)-(28), followed by conversion of electrons $x_{el}$ to gray levels $x$.

Examples of $P(x|k)$ are shown in Fig. 2 for $k = 0$, 1, and 2 input photoelectrons, for the camera parameters listed in Table I. Gray levels above threshold—the dotted vertical line at $x = 210$ gl—contribute a signal proportional to the shaded area under the curve, which gives $P(x > T|k)$. For $k = 0$, this represents the electronic noise probability $p_{el}$, which here is 0.016. For $k = 1$, it gives the probability of getting a "click" from an absorbed photon. This is an effective quantum efficiency $P(x > T|1) = \eta_{EMCCD}$, which here has a value of 0.61.

EMCCDs with sufficiently high gain and low read noise may operate in photon-counting mode and be approximated as an array of single-photon counters. This is the regime in which the above analysis is applicable.

## III. EXPERIMENTAL RESULTS

We compare our theoretical results with experimental measurements of spatially entangled photon pairs using an EMCCD camera. Biphotons are generated via collinear type-I SPDC in a BBO crystal pumped by a spatially filtered 400 nm cw laser diode, and the far field is projected onto an EMCCD camera (Andor, iXon Ultra 897) [see Fig. 3(a)]. The EMCCD consists of a 512×512 array of 16×16 μm² pixels, and is operated at −85 °C (maintained by water cooling), 17 MHz readout rate, with 0.3 μs vertical shift time, and vertical clock voltage of +4 V above default. A 101×101 pixel region of interest centered on the intensity distribution is selected, and the exposure time is fixed to 5 ms. The electronic noise $P(x|0)$ is measured by obtaining a histogram of gray levels from $10^4$ frames collected with the shutter closed [Fig. 3(b)]. This is fit with Eqs. (25)-(28) to characterize the EMCCD; resulting parameters are given in Table I.

Measurements of $\Gamma_{ij}$ are performed at many values of $\langle C \rangle$. The mean count rate is varied by adjusting the attenuation of the pump laser with a continuously variable ND filter. For each mean count rate, $10^4$ gray level images are collected, thresholded at $T = 210$ gl—the value which maximizes the SNR—and processed. A region with uniform singles count rate (uniform irradiance) is selected [Fig. 3(c)], from which $\Gamma_{ij}$ is calculated via Eq. (15) (with $\alpha$ set to one). Figure 3(d) shows the projection of $\Gamma_{ij}$ onto the sum coordinates, $(\boldsymbol{\rho}_1 + \boldsymbol{\rho}_2)/\sqrt{2}$, where the strong peak in the center indicates anti-correlation of the entangled photon pairs. To determine the SNR, we fit the correlation peak to a 2D Gaussian and take its area as the signal; the noise is given by the standard deviation of the background far from the peak. Defining the signal and noise in this way essentially averages the 4D joint probability distribution $\Gamma_{ij}$ over many pixels. Thus, the uniformity of the irradiance (and $\Gamma_{j|i}$) is important to this metric, as spatial variation complicates the analysis. As shown in Figs. 3(e) and 3(f), measurements of the signal, noise, and SNR agree well with theory

Further evaluation of the theory over a large range of $p_{el}$ and $\eta_{EMCCD}$ can be performed by imaging at different threshold levels. For thresholds below the original $T = 210$ gl, both $p_{el}$ and $\eta_{EMCCD}$ increase, as a larger portion of $P(x|k)$ is above threshold. Figure 3(f) shows measurements of the SNR at thresholds from 147 gl to 210 gl (corresponding to $\sigma_{read}$ below and $1.9\sigma_{read}$ above $\mu_{read}$), with good agreement with theoretical fits of Eq. (22) over the entire range.

For higher thresholds, the approximation of the EMCCD camera as an SPCM breaks down. Figure 3(g) shows that measurements of the SNR for thresholds between 233 gl and 388 gl (corresponding to $2.5\sigma_{read}$ and $10\sigma_{read}$ above $\mu_{read}$), disagree with theory. For large $\langle C \rangle$, the measured SNR is much greater than predicted, particularly for the highest thresholds. In this regime, counts are more likely to originate from more than one input electron, $k > 1$. The reason for this, as discussed in section IV below, is that the probably of registering a click for more than one input photon scales differently for an EMCCD than for an SPCM.

We also compare the optimum count rate $\langle C \rangle_{\text{opt}}$ and corresponding maximum SNR found from experiment with theory. Figure 4(a) shows that the measured $\langle C \rangle_{\text{opt}}$ agrees well with Eq. (23) for $p_{el} > 0.01$ (smaller values of $p_{el}$ correspond to higher thresholds, where the model breaks down). Over the range of validity, we may incorporate the threshold dependence of $p_{el}(T) = P(x > T|0)$ [see Fig. 2] and the optimum count rate [Eq. (23)] into the expression for the SNR [Eq. (22)] to predict the maximum achievable SNR as a function of threshold. This curve is plotted in Fig. 4(b), which shows a peak at $T = 210$ gl (thus our preferred operating value). As before, agreement with experiment is very good for all but the highest thresholds.

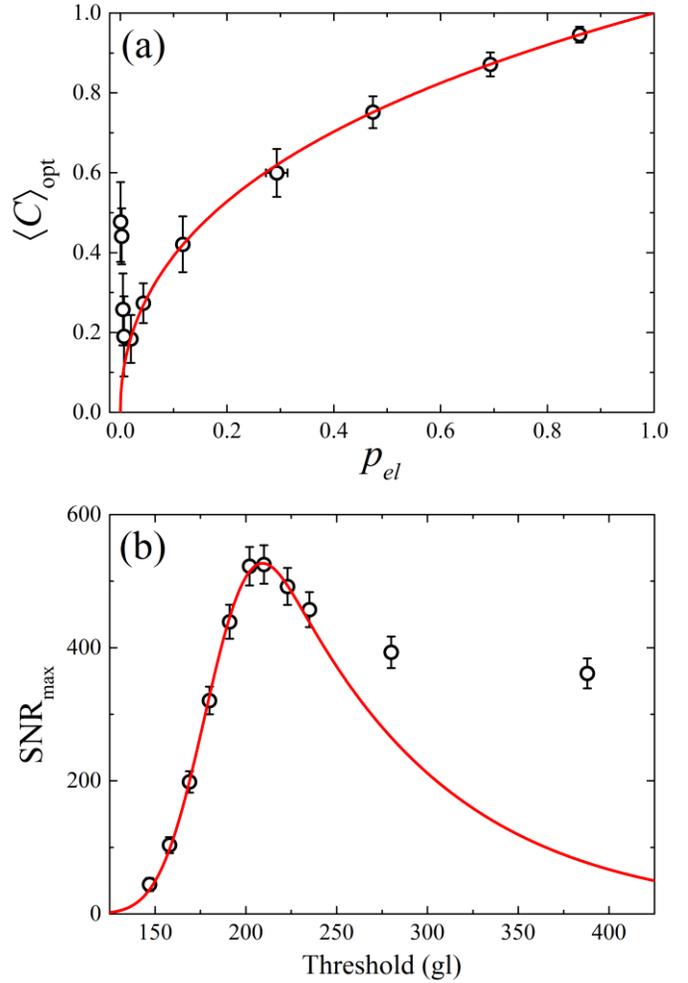

Fig. 4 Comparison of (circles) measured optimum count rate $\langle C \rangle_{\text{opt}}$ and corresponding maximum SNR (SNR$_{\text{max}}$) with (curves) theory. Curve in (a) is $\langle C \rangle_{\text{opt}}$ from Eq. (23) plotted versus $p_{el}$, which shows good agreement for $p_{el} > 0.01$. In (b) the maximum SNR is plotted versus gray-level threshold. The red curve shows Eq. (22) with $\langle C \rangle = \langle C \rangle_{\text{opt}}$ and the known dependence of $p_{el}$ on threshold, i.e., $P(x > T|0)$. There is agreement between theory and experiment for all but the highest thresholds, which correspond to the lowest values $p_{el}$.

## IV. DISCUSSION

By characterizing noise properties of the EMCCD camera, that is, measuring $P(x|0)$, the optimum operating parameters can be deduced. Using Eqs. (22)-(24), the threshold $T$ and count rate $\langle C \rangle_{\text{opt}}$ that maximize the SNR can be found. We have experimentally validated the theory for threshold values within several standard deviations of $\mu_{read}$. Fortunately, the global optimum of the SNR is found in this range, which is therefore where measurements of the biphoton joint probability distribution should be made.

The breakdown of the theory at high thresholds arises from differences between EMCCDs and SPCMs, i.e., how the probability of registering a "click" depends on the number of incident photons. SPCMs are Geiger-mode avalanche devices, whose output is either zero or one depending on whether or not an avalanche was triggered. This results in an avalanche probability that scales with the number of incident photons $n$ as

$$P(1|n) = 1 - (1-\eta)^n. \quad (29)$$

This form of the "click" probability allows the simple insertion of the quantum efficiency as in Eqs. (4) [25]. Applying this concept to the EMCCD camera requires the probability of getting a gray level above threshold to scale as

$$P(x > T|k) = 1 - (1-\eta_{EMCCD})^k. \quad (30)$$

where $\eta_{EMCCD} = P(x > T|1)$ (see Appendix B).

However, EMCCDs do not have the same form of scaling with incident photon number. Figure 5 shows $P(x > T|k)$, calculated via Eqs. (24)-(28) for the EMCCD parameters in Table I. Even for two input photoelectrons, $P(x > T|2)$ (dotted blue curve) is significantly different from $1 - (1 - P(x > T|1))^2$ (dot-dashed maroon curve). This discrepancy grows with both increasing threshold and increasing photoelectron number. For sufficiently low threshold, the approximation of an EMCCD as an SPCM, Eq. (30), is valid. (It even improves with decreasing threshold since $P(x > T|k) \rightarrow 1$.) This explains the agreement between experiment and theory for $T \leq 210$ gl [see Fig. 3(f) and 4]. However, for higher $T$ this approximation becomes incorrect. Because $P(x > T|k) > 1 - (1-\eta_{EMCCD})^k$, the measured SNR is greater than the theory predicts for high thresholds and count rates, as most counts originate from multiple input photoelectrons per pixel.

To further confirm the origin of the discrepancy, we perform numerical simulations using both SPCM and realistic EMCCD responses, i.e., $P(x > T|k)$. Briefly, a Poissonian distribution of photon pairs with mean $\bar{m}$ is sampled for each of $10^6$ frames. The pairs then arrive at the detector per an ideally anti-correlated biphoton joint probability distribution, $\Gamma_{ij} = \delta_{i,-i}$. In each pixel, photons are detected with quantum efficiency $\eta$. For the EMCCD, the gray level at the output is calculated by sampling $P(x|k)$, with the appropriate $k$, and then thresholded. For SPCM simulations, Poissonian noise is added with mean $p_{el}$. For both detector systems, simulated measurements of $\Gamma_{ij}$ are calculated via Eq. (15) (with $\alpha = 1$), from which the SNR is found. This is repeated for many values of $\bar{m}$ to span the entire range of $\langle C \rangle$ from $p_{el}$ to 1.

Simulations of the EMCCD were performed using the parameters in in Table I to model $P(x|k)$ at $T = 280$ gl, which shows excellent agreement with experiment [Fig. 5(b)]. A global scaling factor is applied to the simulations to match the amplitude with experiment. This accounts for differences in unknown quantum efficiency and mean photon number in the experiment, as well as the lower number of pixels used in simulation for computational speed. The remaining deviations from experiment may be due to slight non-uniformity of $P(x|k)$ across the pixels in the frame, inaccuracies in the model [23,24], or fitting errors. Simulations of SPCM's with the same $p_{el}$ were then performed, and match well with our theory. We therefore conclude that the discrepancy between theory and experiment is due to the non-SPCM-like behavior of EMCCDs at high thresholds. A complete characterization of this behavior can be understood my taking into account the full properties of the camera [25].

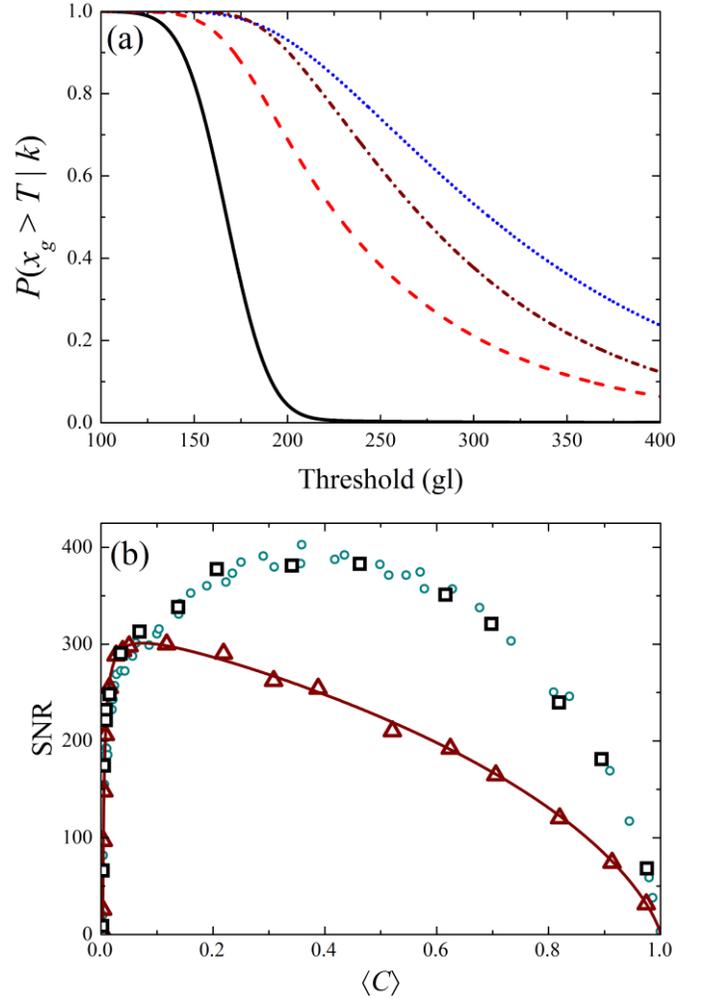

Fig. 5 Discrepancy between thresholded EMCCD and SPCM. (a) Probability of a gray level above threshold, $P(x_g > T|k)$, given (solid black) $k$ = zero, (dashed red) one, and (dotted blue) two input photoelectrons. Curves were calculated with Eqs. (24)-(28) with EMCCD parameters in Table I. Dot-dashed maroon curve is $1 - (1 - P(x > T|1))^2$ which is implicitly assumed in the model. The difference between dashed blue and dot-dashed maroon curves, and those for larger $k$, is the cause of the discrepancy between experimental results at high threshold and theory. (b) Comparison of simulations of (black squares) EMCCD and (maroon triangles) SPCM with (teal circles) experiment and (curve) theory. Simulation parameters were the same as experiment with EMCCD at $T = 280$ gl and SPCM with the same value of $p_{el} = 0.002$, and scaled to the same amplitude as experiment.

## V. CONCLUSION

We have provided a general analytical expression for the SNR for measurements of entangled photon pairs. This expression assumes only a Poissonian distribution of photon pairs and is valid for the full range of count rates up to saturation. There is an optimum count rate at which the SNR is maximized that depends only on the detector noise properties, and may therefore be specified ahead of any quantum experiments. The theory works particularly well for EMCCDs at low thresholds, while for high thresholds the cameras deviate from ideal binary photon counters. These differences are negligible for EMCCDs with low readout noise and high gain when operated with the appropriate threshold. Indeed, the optimum threshold occurs well within the region of validity, even for relatively high read noise, when operated at maximum readout rate [14,24,31]. The SNR curve around the peak is relatively broad, with a falloff for non-ideal parameters that is relatively slow. The results therefore suggest a large operating window for collecting data at significantly higher count rates than is typically done.

## ACKNOWLEDGMENT

This work was supported by DARPA grant HR0011-12-C-0027, and AFOSR grants FA9550-14-1-0177 and FA9550-12-1-0054.

## APPENDIX A: Non-uniform illumination

For non-uniform illumination, the signal is related by

$$\Gamma_{ij} = -\alpha\eta\Gamma_{j|i}\ln\left(\frac{1-\langle C_i\rangle}{1-p_{el}}\right) = -\alpha\eta\Gamma_{i|j}\ln\left(\frac{1-\langle C_j\rangle}{1-p_{el}}\right). \quad (A1)$$

due to the symmetry of the biphoton joint probability distribution. The standard deviation where $\Gamma_{ij} = 0$, i.e., where $\langle C_{ij}\rangle = \langle C_i\rangle\langle C_j\rangle$, is

$$\sigma_{\Gamma_{ij}} = \frac{\alpha}{\sqrt{N}}\sqrt{\frac{\langle C_i\rangle\langle C_j\rangle}{(1-\langle C_i\rangle)(1-\langle C_j\rangle)}} \quad (A2)$$

The SNR is given then by the ratio of Eq. (16) over Eq. (21)

$$\begin{aligned}\text{SNR} &= -\eta\Gamma_{j|i}\sqrt{N}\sqrt{\frac{(1-\langle C_i\rangle)(1-\langle C_j\rangle)}{\langle C_i\rangle\langle C_j\rangle}}\ln\left(\frac{1-\langle C_i\rangle}{1-p_{el}}\right)\\ &= -\eta\Gamma_{i|j}\sqrt{N}\sqrt{\frac{(1-\langle C_i\rangle)(1-\langle C_j\rangle)}{\langle C_i\rangle\langle C_j\rangle}}\ln\left(\frac{1-\langle C_j\rangle}{1-p_{el}}\right).\end{aligned} \quad (A3)$$

In the limit of low count rate, this reduces to

$$\text{SNR} \approx \eta\Gamma_{j|i}\sqrt{N}\frac{\langle C_i\rangle - p_{el}}{\sqrt{\langle C_i\rangle\langle C_j\rangle}} = \eta\Gamma_{i|j}\sqrt{N}\frac{\langle C_j\rangle - p_{el}}{\sqrt{\langle C_i\rangle\langle C_j\rangle}}. \quad (A4)$$

## APPENDIX B: SPCM-like scaling of EMCCD

For a detector with general gray-scale response to incident photons, the conditional probability of a particular gray level $x$ on the number of incidence photons $n$ is given by [25]

$$P(x > T|n) = \sum_{k=0}^{n} P(x > T|k)P(k|n), \quad (B1)$$

where $k$ is the number of generated photoelectrons and

$$P(k|n) = \binom{n}{k}\eta^k(1-\eta)^k, \quad (B2)$$

where $\eta$ is the quantum efficiency (absorption probability). Let $P(x|k)$ be the probability of generating gray level $x$ given $k$ input photoelectrons. If

$$P(x > T|k) = 1 - (1-\eta_{EMCCD})^k, \quad (B3)$$

then Eqs. (B1)-(B3) give

$$P(x_g > T|n) = 1 - (1-\eta\eta_{EMCCD})^n. \quad (B4)$$